\documentclass[10pt,a4paper]{article}
\usepackage[utf8]{inputenc}
\usepackage[english]{babel}

\usepackage{amsmath}
\usepackage{amsfonts}
\usepackage{amssymb}
\usepackage{adjustbox}
\usepackage{hyperref}
\usepackage{comment}
\usepackage{pdflscape}
\usepackage[font=small,skip=0pt]{caption}
\usepackage{hyperref}
\usepackage{xcolor}
\usepackage{comment}

\usepackage{algorithm}
\usepackage[noend]{algpseudocode}

\definecolor{crix}{HTML}{98BF8D}
\definecolor{etf}{HTML}{AE7590}
\definecolor{eth}{HTML}{55BCC2}
\definecolor{btc}{HTML}{EA8E84} 
\definecolor{trading_fees}{HTML}{3E0751}
\definecolor{a2}{HTML}{B0C1CE}
\definecolor{a5}{HTML}{7793AC}
\definecolor{a10}{HTML}{3F678B}

\usepackage{graphbox}
\usepackage{subcaption}

\graphicspath{{R_repo_ETF_on_CRIX/}}

\usepackage{natbib}
\title{ETF construction on CRIX}

\author{Konstantin H\"ausler \footnote{Corresponding author,  Email: konstantin.haeusler@hu-berlin.de. IRTG 1792, School of Business and Economics, Humboldt-Universit\"at zu Berlin, Dorotheenstr. 1, 10117 Berlin, Germany.} \ \ Wolfgang Härdle \footnote{W.I.S.E. - Wang Yanan Institute for Studies in Economics, Xiamen University, Xiamen, 361005, Fujian, China. C.A.S.E. - Center for Applied Statistics and Economics, Humboldt-Universität zu Berlin, Unter den Linden 6, 10099 Berlin, Germany. Singapore Management University, Singapore. Department of Mathematics and Physics, Charles University Prague, Ke Karlovu 2027/3, 12116 Praha 2, Czech. National Yang Ming Chiao Tung University (NYCU), No. 1001, Daxue Rd. East Dist., Hsinchu City 300093, Taiwan. \vspace{0.5cm} \newline Financial support from the Deutsche Forschungsgemeinschaft via the IRTG 1792 "High-dimensional, Non-stationary Time Series", Humboldt-Universit\"at zu Berlin, is gratefully acknowledged. This research has also received funding from the European Union's "FIN-TECH: A Financial supervision and Technology compliance training programme" under the grant agreement No 825215.}}
\begin{document}
\maketitle

\section*{Abstract}

Investments in cryptocurrencies (CCs) remain risky due to high volatility. Exchange Traded Funds (ETFs) are a suitable tool to diversify risk and to benefit from the growth of the whole CC sector. We construct an ETF on the CRIX, the CRyptocurrency IndeX that maps the non-stationary CC dynamics closely by adapting its constituents weights dynamically. The scenario analysis considers the fee schedules of regulated CC exchanges, spreads obtained from high-frequency order book data, and models capital deposits to the ETF stochastically. The analysis yields valuable insights into the mechanisms, costs and risks of this new financial product: i) although the composition of the CRIX ETF changes frequently (from 5 to 30 constituents), it remains robust in its core, as the weights of Bitcoin (BTC) and Ethereum (ETH) are robust over time,
ii) on average, a portion of 5.2\% needed to be rebalanced at the rebalancing dates, iii) trading costs are low compared to traditional assets,
iv) the liquidity of the CC sector has increased significantly during the analysis period, spreads occur especially for altcoins and increase by the size of the transactions. But since BTC and ETH are most affected by rebalancing, the cost of spreads remains limited. 
\\
\\
\textbf{Keywords:} Cryptocurrency, ETF, CRIX, Market Dynamics 
\\
\textbf{JEL Classification:} G19

\newpage
\section{Introduction}{\label{intro}}

Investments in digital assets remain risky due to high volatility. A suitable tool to benefit from the growth of this emerging sector while reducing risk through diversification are exchange traded funds (ETF). We construct an ETF on the CRIX (cf. \cite{davis2016}, \cite{trimborn2018}), an index for the cryptocurrency (CC) sector that is listed by S\&P Global. In doing so, we reveal the underlying mechanisms, risks and costs of such a new financial product and pave the way for its implementation. \\

The CC sector is still characterized by high volatility (cf. \cite{haerdle2020}), but there are early signs that it is consolidating, e.g. the level of regulation is increasing and CCs are becoming more integrated into the financial sector. A suitable instrument to benefit from the growth of the entire sector and to avoid idiosyncratic risks of individual CCs are ETFs. So far, there exist several CC-indices on which ETFs can be built, for further details refer to \cite{haeusler2022}. Among them, the CRIX, a CRyptocurrency IndeX developed by \cite{trimborn2018} at the Blockchain Research Center at Humboldt Universit\"at zu Berlin \href{https://blockchain-research-center.com}{[Link]}, is characterized by high accuracy in mapping the dynamics of the CC sector. Accuracy is achieved by dynamically adjusting the number of constituents and their weights, in order to reflect the dynamics of the entire market with as few CCs as possible. Its index data is publicly available at \href{royalton-crix.com}{royalton-crix.com}. In this article, we construct a CRIX ETF to understand the underlying mechanisms and identify potential risks and costs. \\

For the following analysis, a deep understanding of the functioning of the CRIX and its rebalancing mechanism is crucial. Therefore, in Section \ref{sec::data}, the methodology of the CRIX as outlined in \cite{trimborn2018} is introduced. Descriptive statistics provide some initial insights into the topic. Then the individual components of the scenario analysis will be explained. Since we are primarily interested in the operation and risks of issuing a CRIX ETF, we focus on the practical aspects of ETF issuance. First of all, an appropriate CC exchange is needed to execute the trades. A brief discussion of appropriate exchanges and their fee schedules is presented in Section \ref{sec::cost_types}. In addition to the trading fees, spreads are also incurred. An analysis of the spread structure and the liquidity of each CC follow in Section \ref{sec:costs_rebalancing}. We approximate bid-ask-spreads that might emerge at rebalancing by historical  high-frequency trading data snapshots from Binance. Even though Binance is not an eligible exchange for institutional investors, its API allows to obtain data on spreads. And as the liquidity of Binance is among the highest (cf. \cite{brauneis2021}), it allows to estimate a lower bound for spreads. Throughout the period of analysis, we model capital in-\& outflows stochastically from/to the ETF by a cumulative random walk that is driven by the Google Trends search results for "CC ETF", an indicator for how much attention CC ETFs attract. 

Then, the scenario analysis of the ETF for the period of 07-2020 to 06-2021 is carried out in the following steps: Initially, an hypothetical amount of capital is allocated among the CRIX constituents, relative to their respective weights. Periodically at the rebalancing dates, CCs are bought/ sold at their current market prices according to their updated CRIX weights. At each step, an analysis of the composition of the ETF portfolio, its rebalancing quantities, as well as the incurred costs (consisting of trading fees and spreads) is conducted. Several checks are conducted to ensure that everything went right. \\

In the process, we gain insights into various topics: How robust are CRIX constituents and how does this affect the portion of the portfolio that needs to be rebalanced at each rebalancing date? 
How do the spreads depend on the specific CC, the CC exchange and the trading volume, and how do they evolve over time?
As the CC sector is not yet consolidated, does the emergence and disappearance of CCs lead to high fluctuations in the index constituents, and how is this reflected in the rebalancing costs? 
\\
\\
The scenario analysis yields various insights: First, the composition of the CRIX ETF varies throughout the period of analysis widely, from 5 to 30 constituents. Even though the number of constituents varies a lot, the share of the portfolio that had to be rebalanced amounts on average to $5.2 \%$ (minimum $2.2 \%$ in August 2020, maximum $15.2 \%$ in June 2021). However, large shares of the index do not need to be rebalanced, as Bitcoin (BTC) and Ethereum (ETH) reveal large and robust weigths throughout the period of analysis. The number of constituents is not correlated to the portion of the portfolio that needs to be rebalanced.\\

Second, the CRIX ETF seems very diversified, as it covers up to thirty CCs. However, diversification is limited, as  BTC and ETH dominate the index: their weights sum up to roughly 80 \% of the portfolio, and this share is robust over the period of analysis. In general, diversification in the CC sector is difficult, as \cite{keilbar2021} identified cointegration relationships among the top CCs by market capitalization. We recommend limiting the number of constituents, because \cite{haeusler2022} revealed that a bigger set of constituents does not automatically lead to a more accurate mapping of the dynamics of the CC sector. Reducing the number of constituents can also simplify the management of the ETF. \\

Third, BTC and ETH are the constituents that are most rebalanced. Since their liquidity is the highest among all CCs, their spreads are relatively low, which in turn limits trading costs. However, the frequent in- and exclusion of BTC and ETH  causes recurring rebalancing costs. We suggest to alter the CRIX ETF algorithm to optimize the rebalancing quantities. \\

Fourth, the issuance of a CRIX ETF involves trading costs, which consist of trading fees and spreads. Even though trading fees are low (expected 0.2\%, conditional on the exchange and transaction volume), spreads can cause considerable costs, as the weights of up to 30 CRIX constituents need to be updated and not all CCs are as liquid as BTC. Overall, spreads decrease over the analysis period. Depending on the assumptions on the structure of spreads, we estimate the median of the spreads at the rebalancing dates between 0.03\% and 0.27\%. \\

Fifth, the performance of the CRIX ETF is difficult to assess econometrically, as the CC sector is highly non-stationary and thus the performance of the ETF depends on the time interval under review. As a benchmark, we compare the performance of the CRIX ETF with a simple investment in BTC. Interestingly, the Sharpe Ratio of the ETF outperforms BTC.   \\

Some issues related to the issuance of a CC ETF are not addressed here for feasibility reasons, e.g., synthetic construction of CC ETFs (as the market for derivatives is not yet sufficiently liquid) or an analysis of market depth, and the extent to which the execution of large rebalancing positions can lead to market movements. Section \ref{sec::conclusion} addresses such issues and concludes with practical comments and topics for further research.

\section{Data}{\label{sec::data}}

For the scenario analysis of a CC ETF, data is obtained from three sources: firstly, the CRIX is used as the underlying index on which the CC ETF is constructed. Secondly, prices of CCs are obtained from coingecko.com, a platform that collects and processes pricing and trading data of CCs from a large range of CC exchanges. The authors thankfully acknowlegde their freely accessible API. Thirdly, order book data of several CC exchanges is obtained from the Blockchain Research Center \href{https://blockchain-research-center.com}{[Link]} at Humboldt Universität zu Berlin. 

\subsection*{The CRIX - a CRyptocurrency IndeX}

\begin{figure}[h!]
\centering
\includegraphics[width=\textwidth]{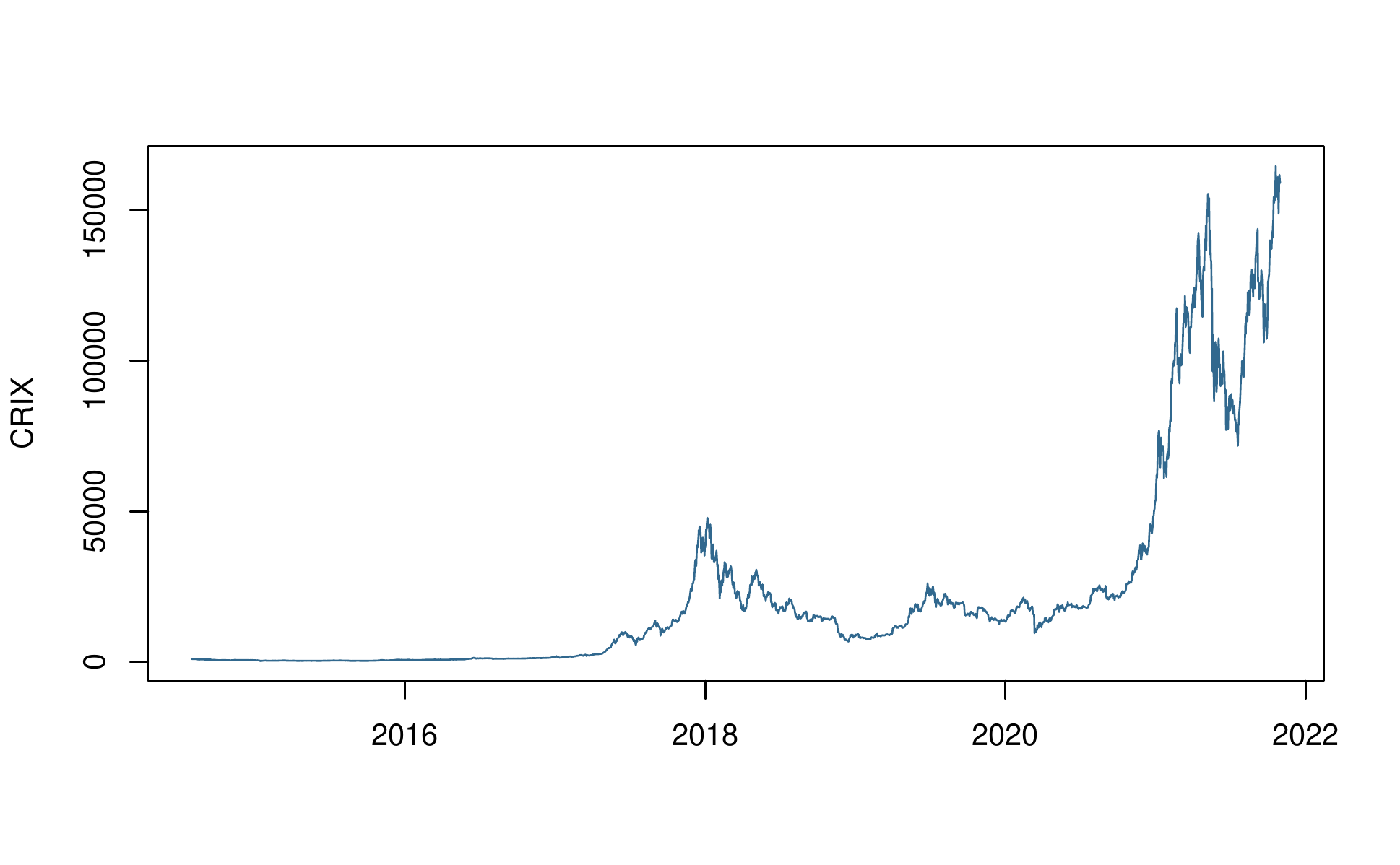}
\caption{CRIX, a CRyptocurrency IndeX. Data source: \href{royalton-crix.com}{royalton-crix.com} 
}
\label{graph::CRIX_plot}
\end{figure}

The CRIX (see Figure \ref{graph::CRIX_plot}) serves as the underlying CC index on which the ETF will be built. Among the existing CC indices, the CRIX (developed at the Blockchain Research Center by  \cite{davis2016}) is convincing because it optimally solves the fundamental trade-off faced by any index, an accurate mapping of the market dynamics at a sparse number of constituents. By adjusting the number of constituents dynamically, the CRIX ensures high accuracy in representing the CC market dynamics (cf. \cite{haeusler2022}). The dynamic adaption to the market is important, because the CC sector is not yet consolidated and under steady transition. The CRIX is constructed as a Laspeyres index that weights each constituent $i$ by its market cap:

\begin{align*}
CRIX_t = \frac{\sum_i P_{it}Q_{i0}}{\sum_i P_{i0}Q_{i0}}
\end{align*}

where $P_{it}$ refers to the price of constituent $i$ at time $t$, and $Q_{i0}$ the amount of constituent $i$ at time point $0$. Every 3 months, the optimal set of constituents is determined in the following way: the log-returns of several CC portfolios $CCP$, $\varepsilon(k)_t^{CCP}$, consisting of $k=1,2,3,..., K$ CCs (sorted from top to down by market capitalization), and the log-return of the total market $TM$, $\varepsilon(K)_t^{TM}$, are computed. The goal of the CRIX is to choose $k$ such that the distance $\lVert \hat{\varepsilon}(k) \rVert^2 = \lVert \varepsilon(K)^{TM} - \varepsilon(k)^{CCP}\rVert^2$ is optimized at a sparse number of CCs. This is achieved by the AIC:

\begin{align*}
    AIC\{\hat{\varepsilon}(k), k\} =  -\log L\{\hat{\varepsilon}(k)\} + 2k
\end{align*}

where $L$ is the Likelihood function that maximises the density of $\hat{\varepsilon}(k)_t$ over all time periods $t$. 
Thus, AIC minimizes the distance of the CRIX portfolio to the total market, but penalizes for an increasing number of CCs. Thereby the previously mentioned trade-off (accuracy vs. a sparse number of constituents) is solved optimally.\\

Important for the further analysis of the ETF is that the weight of each constituent is updated monthly according to its market capitalization. In addition, the number of constituents is updated every three months, i.e. some constituents can be dropped/ included. 

\begin{figure}[h!]
\centering
\includegraphics[width=0.8\textwidth]{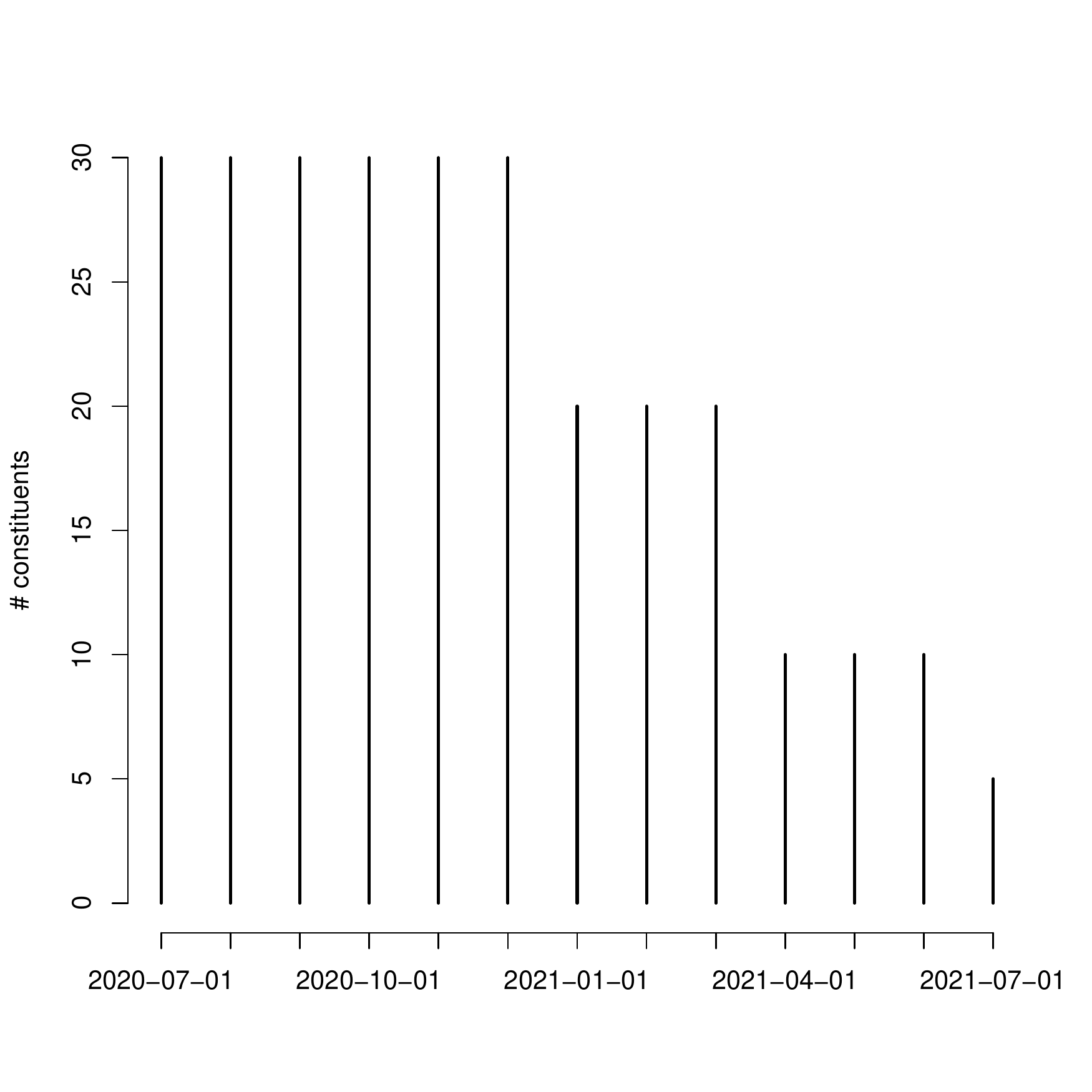}
\caption{Number of CRIX constituents over time. The set of constituents is updated every quarter, i.e. on the 1st of Jan/ Apr/ Jul/ Oct. The weight of each constituent is updated monthly. 
}
\label{graph::crix_constituents}
\end{figure}

For the period of analysis, the set of constituents varied between five and thirty, see Figure \ref{graph::crix_constituents}. Each bar in the chart refers to the number of CRIX constituents at each month. The fluctuation in the composition of the CRIX is high, because the CC sector is still rising and under steady transitions. Over the period of analysis, the number of constituents decreases.  The variation in the number of CRIX constituents supports the choice of the CRIX as the underlying CC index for the ETF: due to the volatility of the CC sector, it is not sufficient to set a fixed set of CCs to represent the entire CC sector. The variation of the number of constituents indicates that a dynamic adjustment is necessary to reflect the overall market dynamics.

\begin{figure}[h!]
\includegraphics[width=\textwidth]{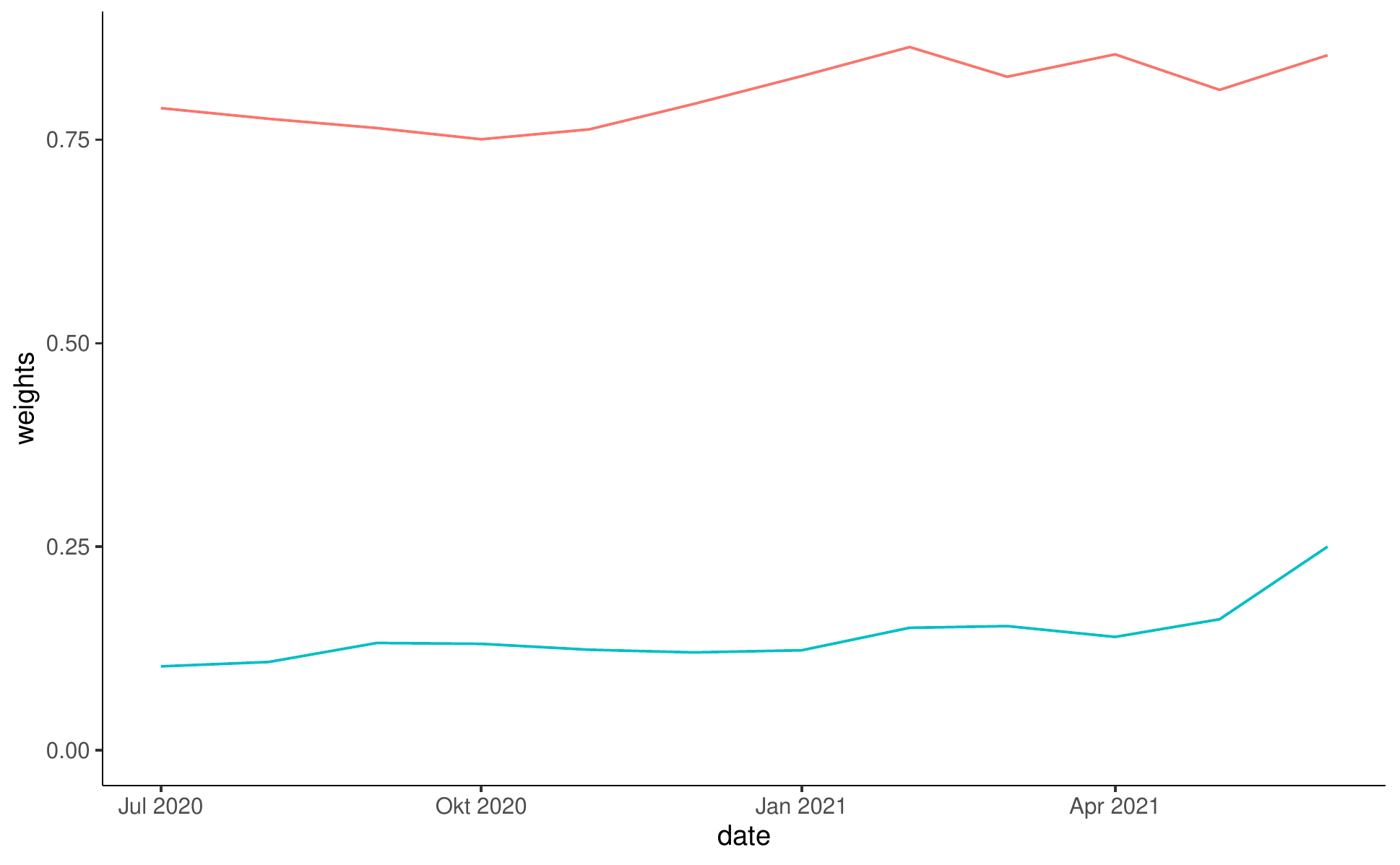}
\caption{Stacked CRIX weights of {\color{btc}BTC} and {\color{eth}ETH} over time. Weights vary little, even though the number of constituents does. 
}
\label{graph:ETH_BTC}
\end{figure}

As the constituents are weighted by their market capitalization, BTC and ETH reveal the biggest weights and dominate the index, cf. Figure \ref{graph:ETH_BTC}. Interestingly, BTC and ETH represent roughly 70\% - 80\% of the constituents weights, and this dominance persists over time. This observation could have two effects: first, it is very likely that the dominance of BTC and ETH makes the core of the ETF very robust. On the other hand, the dominance of these two CCs may lead to a loss of diversification, one of the reasons why ETFs are issued.

\section{Methodology}\label{sec::cost_types} 

In this section we briefly outline the elements of the ETF construction. The elements cover the entire process of issuing an ETF, from choosing a broker or exchange to modeling capital deposits and withdrawals. The ultimate goal is to understand the rebalancing mechanisms, to identify possible risks and to determine the cost structure of issuing such a CC ETF. \\

We divide the costs into fixed costs and variable costs. Fixed costs comprise any kind of administration or management costs, as well as fees/ commissions for licensing such a financial product. In addition, storage and safekeeping costs for CCs, a relevant point for digital assets (cf. \cite{grobys2020}),  account to fixed costs. As these costs are fixed and company specific, we disregard them in the following analysis. \\

The variable costs consist of rebalancing and adjustment costs. At each rebalancing date (the 1st of each month), the weight of each constituent gets updated, and therefore CCs need to be bought and sold at current market prices. The rebalancing might result in trading costs, which consist of trading fees and spreads. The factors on which these rebalancing costs depend will be analyzed in the next Section \ref{sec:costs_rebalancing}. 
Furthermore, the portfolio positions need to be adjusted due to capital in- and outflows, as customers deposit/ withdraw funds from/ to the ETF. We propose a simple approach to model these cash flows realistically in Section \ref{sec:costs_adjustment}.

\subsection{Rebalancing}\label{sec:costs_rebalancing}

The rebalancing costs depend on several factors. First, trading fees depend on the choice of the broker and the transaction volume. Considerations regarding eligible exchanges follow in the next paragraph. On top of the trading fees, spreads might cause considerable costs. An analysis of the structure of spreads is conducted by examining high-frequency order book data. Ultimately, the rebalancing costs depend on the rebalancing quantities that have to be reallocated at each rebalancing day, i.e. on the robustness of the weights of the CRIX constituents. The results in Section \ref{sec:simulation} will shed light on this topic.


\subsubsection{Appropriate Exchanges}{\label{sec:exchanges}}

To quantify rebalancing costs, an appropriate CC exchange is needed on which trades can be executed and which meets several criteria: it must be in line with the country's legal and regulatory standards, as such funds have to fulfill certain official requirements based on which they are allowed to invest the money of their costumers. Most of the currently existing CC exchanges do not fulfil such legal requirements, and thus cannot be considered as a trading platform for the purpose of issuing an ETF. Besides the legal issues, an appropriate CC exchange must fulfill other requirements. The exchange needs to cover a broad range of CCs, because the number of CRIX constituents varied broadly throughout the period of analysis. Furthermore, the trading volume must be high to guarantee the necessary liquidity to adjust the positions. Finally, among the eligible exchanges, the exchange with the lowest fees is preferable. \\


There are several exchanges and brokers that allow trading CCs, each with several (dis-) advantages: some have high trading volumes but are not regulated (e.g. Binance), others meet regulatory requirements but cannot provide high liquidity or high coverage of CCs. For example, the following exchanges could be shortlisted (without claiming completeness, there may be other appropriate exchanges): Bitstamp, Coinbase and Kraken. Each of them is regulated in the US as well as in Europe, and they all cover a broad (but still limited) number of CCs. The fee schedules of these exchanges are quite similar: the trading fees depend on the transaction volume and the maker/taker side and decrease with increasing volume. The trading volume of Coinbase exceeds that of the other two exchanges.

\begin{figure}[h!]
\centering
\includegraphics[width=0.8\textwidth]{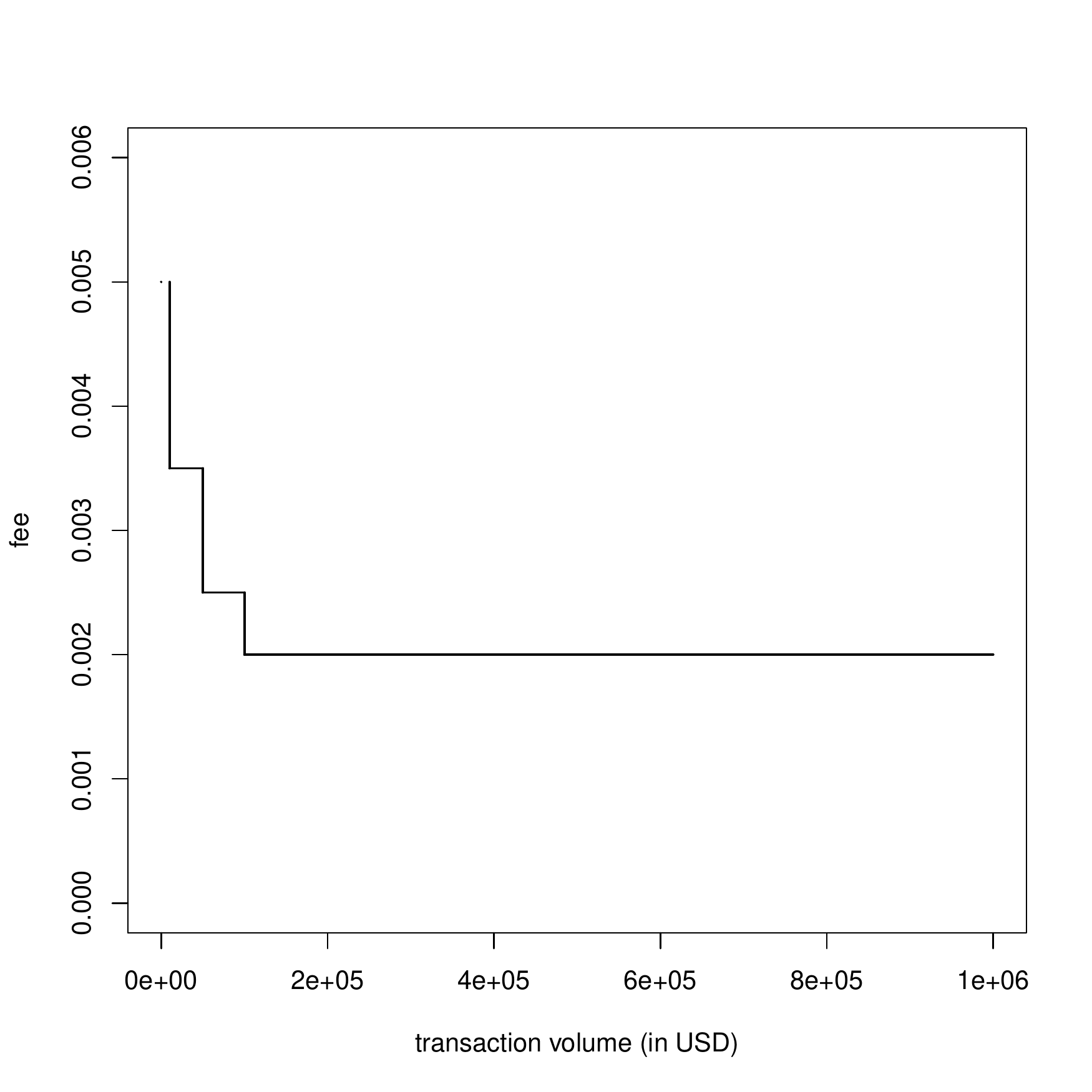}
\caption{Fee schedule similar to the fee schedule of Coinbase Pro. Transaction fees (in $\%$) decrease by increasing transaction volume (in thousand USD). 
}
\label{graph:fee_schedule}
\end{figure}

In the following scenario, we do not commit to a particular exchange, but assume a fee schedule that decreases stepwise as transaction volume increases, similar to the ones of the previously mentioned exchanges. For details of the fee schedule, see Figure \ref{graph:fee_schedule}. \\

\subsubsection{Spreads}{\label{subsection::spreads}}

On top of the trading fees, spreads might cause considerable costs. The structure of spreads and how they depend on the specific CC and their liquidity is the topic of this section. Furthermore, we explore how spreads evolved over the period of analysis and whether the CC sector gained efficiency over time. \\

Bid-ask-spreads emerge whenever there is a difference between the price a buyer is willing to pay and the price a seller is willing to accept. Even though the market capitalization of the CC sector is rising (cf. \cite{burda2021}), it is worth analyzing whether the liquidity of digital assets is yet at a level at which spreads can be disregarded. To quantify to which extent spreads might emerge, we will analyse Bitcoin, the CC with the highest overall trading volume, to obtain a lower bound for the spreads. Order book data for BTC in 2021 is obtained from Binance. Even though Binance can not be considered an appropriate CC exchange (see Section \ref{sec:exchanges}), it offers the highest liquidity among all CC exchanges  (cf. \cite{brauneis2021}). Thus, analyzing the spreads of Bitcoin, the CC with the highest trading volume, at one of the exchanges with the highest liquidity, will yield an estimate for the lower bound of spreads. Unfortunately, we cannot analyze the spreads of all cryptocurrencies individually because we do not have the order book data of all CRIX constituents, and this would be computationally very intense. Therefore, we model the spreads of all other CCs in the next step by scaling their spreads relative to the BTC spreads. We use the 24h trading volume as an indicator of liquidity and use it as a benchmark to scale the spreads of the other CCs over the period of analysis.

\subsubsection*{Inference on BTC spreads}
There exists a broad range of literature that relates spreads to the liquidity of exchanges, market depth and information asymmetries (e.g. earnings news), e.g.  \cite{wang2000trading} identify an inverse relationship between trading volume and bid-ask spreads, \cite{lee1993spreads} show that "wide spreads are accompanied by low depths, and that spreads widen and depths fall in response to bigger volume [...][and] in anticipation of earnings announcements". However, most of the existing literature refers to traditional assets and not to CCs.

Recent articles examine the liquidity of CC exchanges, most of them only for the top CCs by market cap. \cite{dyhrberg2018investible} analyze the liquidity and transaction costs of BTC at several exchanges (Kraken, Gemini, Gdax) by using full order book data of the turn of the year 2017/18. Their analysis shows that the average effective BTC spread is between 5.6 bps and 22.51 bps, lower than the spreads of equity markets. In an influential article, \cite{brauneis2021} compare the liquidity of BTC and ETH of various CC exchanges by several liquidity measures (e.g. the (il-)liquidity measure by \cite{amihud2002}, among others). Their study reveals that the most liquid exchange is Binance and the most liquid CC is BTC. 

Based on their results, we use the spreads of BTC at Binance as a lower bound for the spreads of other CCs. Thus, we investigate the relationship between spreads and order size by analyzing order book data from Binance. The order book data from Binance contains only filled orders, which does not allow to directly infer on spreads. As a workaround, we approximate spreads by taking the difference of prices a buyer or seller faced whenever her order was filled by more than one counterpart. Since we are primarily interested in the spreads at the rebalancing dates, we regress the approximated spreads (in \% of mean trade price) on the transaction volume (in USDT) for each rebalancing date. 

\begin{figure}[h!]
\centering
\includegraphics[width=\textwidth]{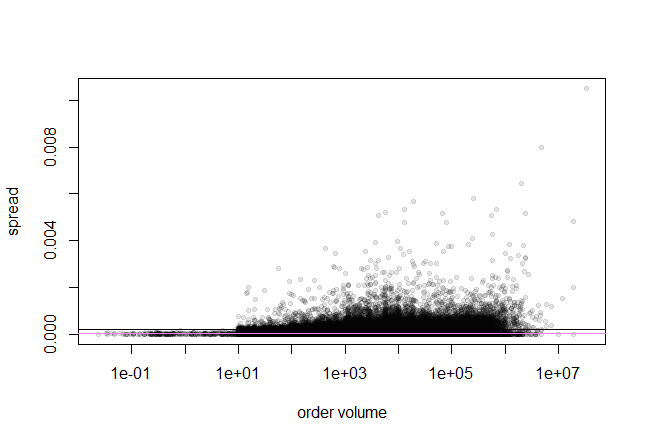}
\caption{Spreads (in \% of transaction price) vs transaction size (in USDT) of Bitcoin order book data from Binance. Violet line: mean regression. Black line: 95\% quantile regression. Number of transactions: 2,365,409. Data source: \href{blockchain-research-center.com}{blockchain-research-center.com}
}
\label{graph::spreads}
\end{figure}

Figure \ref{graph::spreads} illustrates the approximated spreads as a function of the transaction size of BTC aggregated over the rebalancing dates of 2021, which comprises $N=2,365,409$ transactions. 

The approximated spreads vary between 0\% and 1.05\%, on average they are 0.0018\%. Figure \ref{graph::spreads} indicates that the size of spreads increases with the size of the orders. We fit an OLS regression (violet line) and a 95\% quantile regression (black line) to the data. For large transactions (e.g. > 1e06 USDT), there are only very few data points, and in this range the fits differ largely. For smaller transactions though, the fits do not differ much. In the scenario analysis, we take the 95\% quantile instead of the mean regression, as the rebalancing does not occur frequently, and thus no limit theorem is applicable. This means that the empirical costs do not converge to their expected value, because the rebalancing occurs not multiple times, but only at the certain dates. As an issuer of an ETF, one wants to cover these costs with a high probability, we assume 95\%. The estimated coefficients of the spread-volume relationship of BTC are spread = 1.866219e-04      + 5.546762e-09 * volume.



The presence of few data points with high transaction volume makes the analysis imprecise in the upper range. Inference in this domain is therefore difficult. In addition, large transactions can potentially move the market. The results of the following scenario should therefore be treated with caution; they cannot simply be scaled up. 

\subsubsection*{Scaling altcoin spreads relative to BTC spreads}

Unfortunately,  we do not have the historical order book data of other CCs than BTC (i.e. altcoins), and thus, have to approximate their spreads. We do this by taking the 24h trading volume of each CC as a measure of their liquidity, and then scale their spreads relative to the spreads of BTC and its and 24h trading volume around the reference date of 2021-06-01 (the end of our period of analysis, we used data from $\pm 3$ days around the reference date). 

\begin{eqnarray}
spread_{cc,t} = spread_{BTC, 2021} \times \sqrt[a]{\frac{24hTV_{BTC, 2021}}{24hTV_{cc,t}}}
\label{eqn::relBTC}
\end{eqnarray}

\begin{figure}[h!]
\centering
\includegraphics[width=0.76\textwidth]{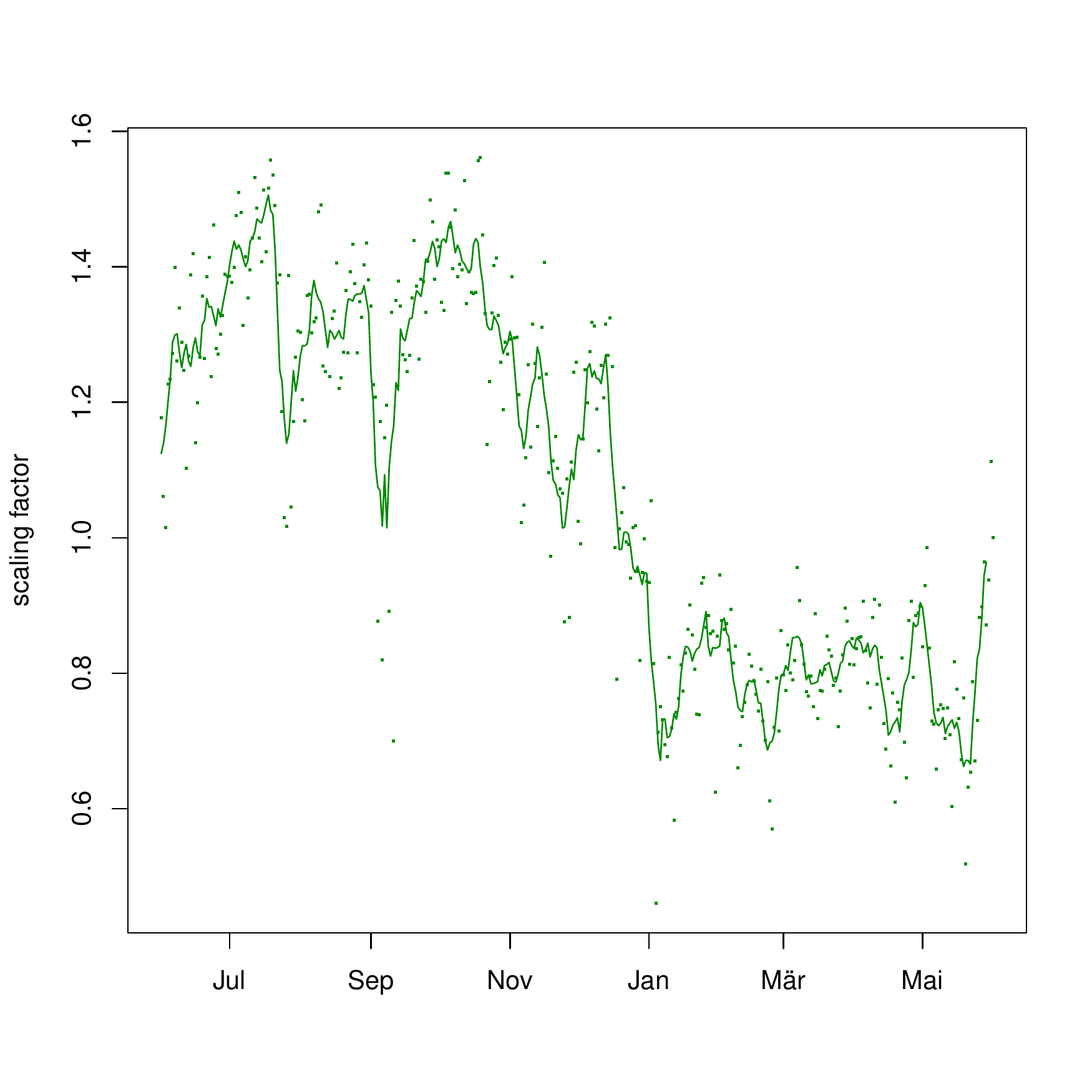}
\caption{Scaling factor of BTC spreads over the entire period of analysis relative to BTC spreads in 2021-06-01. The scaling factor is computed as indicated in Equation \ref{eqn::relBTC}, with $cc=BTC$ and $a = 5$. Daily values (dots) are smoothed by a moving average of one week (solid line). 24h trading volume data obtained from \href{coingecko.com}{coingecko.com}. 
}
\label{graph:relBTC}
\end{figure}

\begin{figure}
\includegraphics[width=\textwidth]{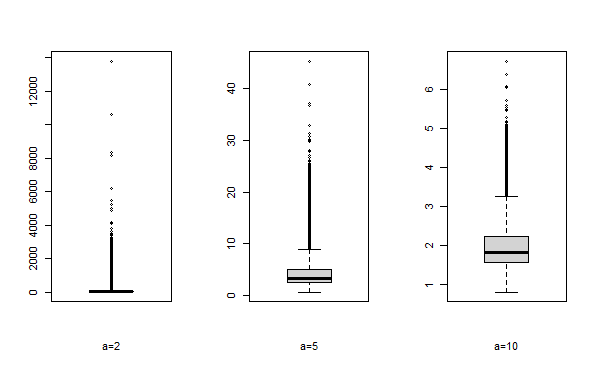}
\caption{Boxplots of scaling factors, i.e. $\sqrt[a]{\frac{24hTV_{BTC, 2021}}{24hTV_{cc,t}}}$, cf. Equation \ref{eqn::relBTC}
over the entire period of analysis, with $a = \{2,5,10\}$.
}
\label{graph:relBTC_boxplots}
\end{figure}

Thus, the spread of CC $cc$ at time $t$ is relative to the spread of $BTC$ at time $t= 2021-06-01$, and scaled by the $a$-th square root of the 24h-trading volume $24hTV$ of BTC in 2021-06-01 relative to the 24h trading volume of the specific CC. Figure \ref{graph:relBTC} illustrates this for the case of BTC itself. Daily values (dots) are smoothed by a moving average of one week (solid line). Interestingly, the liquidity of BTC increased over the period of analysis. Assuming $a=5$, the historical, unobserved spread of BTC in July 2020 is 1.3 times the observed spread of BTC in July 2021. Crucial for the following analysis is the choice of $a$, i.e. how to scale the spreads of any CC other than BTC relative to BTC. Up to the knowledge of the authors, there is no methodological framework to address this problem. Therefore, to investigate the effect of the choice of $a$ on the analysis, we do not set a fixed value for $a$, but choose several grid values: $a = \{2,5, 10\}$. Figure \ref{graph:relBTC_boxplots} displays the distribution of the scaling factors for these grid values of $a$.\\

\subsection{Capital deposits/ withdrawals
}\label{sec:costs_adjustment}

\begin{figure}[h!]
\centering
\includegraphics[width=0.9\textwidth]{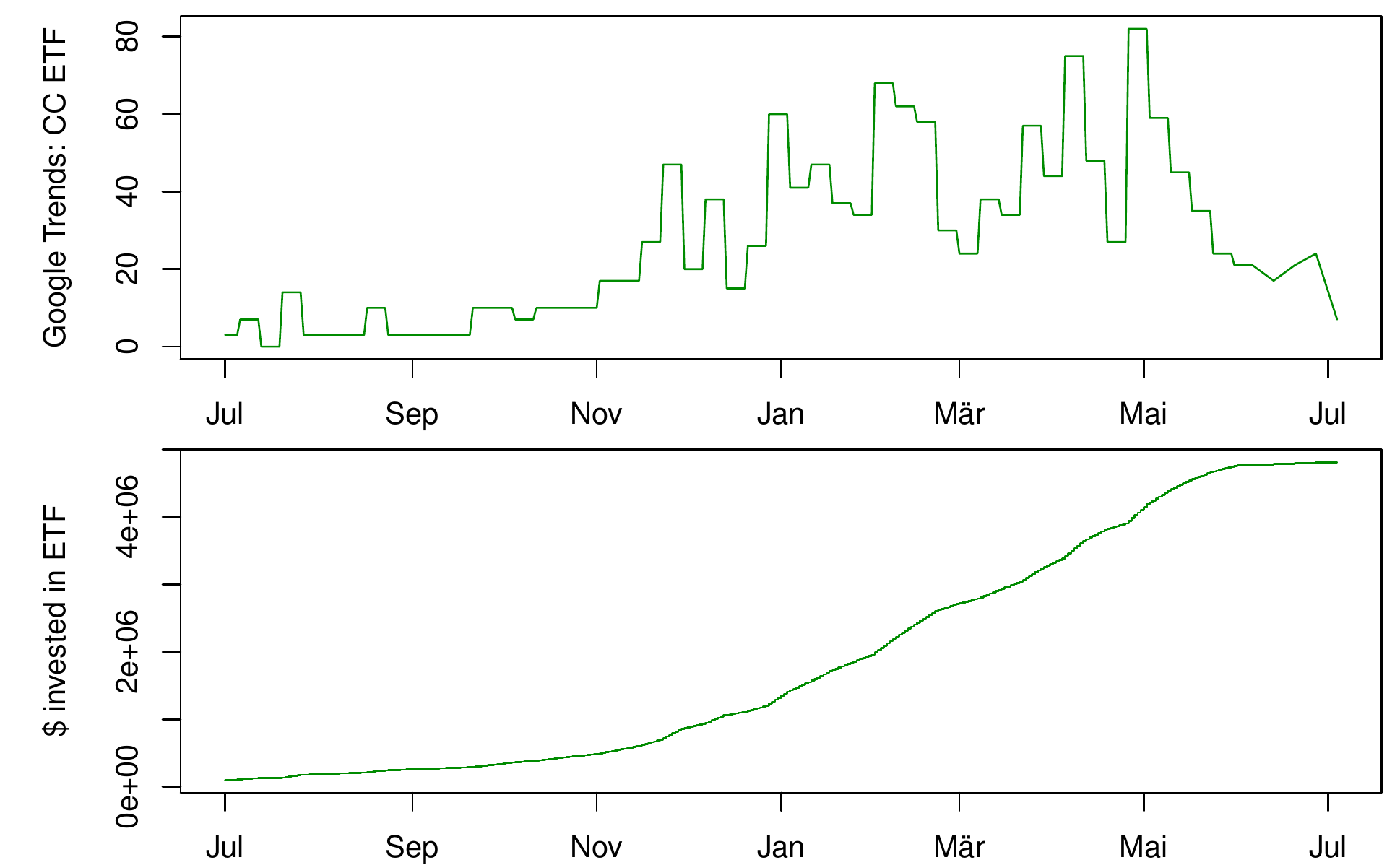}
\caption{Upper chart: Google Trends search for "Cryptocurrency ETF". 2020-07 to 2021-06. Source: \href{ https://trends.google.de/trends/}{Google Trends}. Bottom chart: capital deposits/ withdrawals, modeled by a cummulative random walk, whose increments are asymetrically correlated to the Google Trends time series.
}
\label{graph:gt}
\end{figure}

Capital deposits and withdrawals by customers can result in steady adjustment costs. We assume that the demand and attention for CC ETFs is correlated with Google search requests for "CC ETFs" and thus model capital in- \& outflows through a random walk that is correlated with this time series. The top chart in Figure \ref{graph:gt} shows the time series for "CC ETFs" searched on Google. Interestingly, this time series seems to be correlated to the CRIX: the rise in search queries coincides with the growth of the CRIX at the turn of the year 2020/21, followed by the crash in mid-2021. 

We further assume that ETF investors are rather long-term investors, and therefore capital withdrawals are less frequent. Thus, we model the capital in- \& outflows by a cumulative random walk: increments are a random variable that is asymmetrically correlated with the "attention" measure described above, that is, an increase in searches leads to more capital deposits, but a decrease leads only to a lower capital outflow. The bottom chart of Figure \ref{graph:gt} shows the dynamics of the capital flows: the initial investment starts at 1.000.000 USDT (in general, the capital flows can be scaled arbitrarily), the increases are random but correlated with the Google search queries. For simplicity, deposits/ withdrawals will only be possible at the rebalancing dates (the first week of each month).

\section{ETF construction on CRIX}{\label{sec:simulation}}

In order to understand the mechanisms of issuing an ETF on the CRIX, we run the scenario analysis for the period of 2020-07-01 to 2021-06-01. By doing so, we expect to gain insight into the rebalancing procedure and the costs and risks that may arise when issuing such a financial instrument. Since the CC sector is not yet consolidated and is characterized by high volatility (cf. \cite{haerdle2020}), and the CRIX is an adaptive index that dynamically adjusts to market dynamics, we are particularly interested in how the dynamic adaption affects the rebalancing quantities and the cost structure.
\\

In the scenario analysis, the construction of the CRIX ETF is carried out in the following steps:
At 2020-07-01, the start of the analysis, the initial investment is allocated among the CRIX constituents according to their CRIX weights. At each rebalancing date (in the first week of each month), the value of the portfolio is computed, i.e. the value of each constituent at current prices is calculated. Then the weights of all constituents are updated, and accordingly CCs are bought and sold at current market prices. Furthermore, capital deposits to the ETF (as described in Section \ref{sec:costs_adjustment}) are distributed among the constituents.To ensure that the rebalancing has worked well, we check whether the value of the ETF is proportional to the CRIX time series. In addition, we sum up all trading costs (according to the fee schedule in Figure \ref{graph:fee_schedule} and the spreads in Section \ref{sec:costs_adjustment}). \\

Already in the planning of the scenario analysis, several research questions arise: to what extent is the volatility of the CC sector reflected in the rebalancing quantities and costs? How robust are the weights of the CRIX constituents? What fraction of the portfolio needs to be rebalanced on each rebalancing date? Can we estimate a break-even point regarding the cost structure? Answers to these questions are given in the following section.

\section{Scenario Analysis}\label{sec::results}

The scenario analysis of an ETF on the CRIX yields interesting insights into the functioning of an CC ETF. In the following, we will provide an overview of the findings. 
We start with the weights of the constituents of the CRIX and their robustness, analyze the fractions of the portfolio that need to be updated, and draw conclusions regarding the associated costs. We perform some control checks to make sure that the rebalancing procedure went correctly. An analysis of the performance of the CC ETF is not very informative, since the non-stationarity of the whole sector makes the results strongly dependent on the choice of the period of analysis. As a benchmark, we compare the performance of the CRIX ETF with a simple investment in BTC.

\paragraph{
Robustness of constituent's weights} Due to its still young existence, constant technical innovations and the speculative behavior of investors, the CC sector is under constant transformation and high fluctuation. In order to accurately reflect these dynamics, the CRIX regularly updates the number of index members and their weightings, as described in Section \ref{sec::data}. \\

\begin{figure}[h!]
\includegraphics[width=\textwidth]{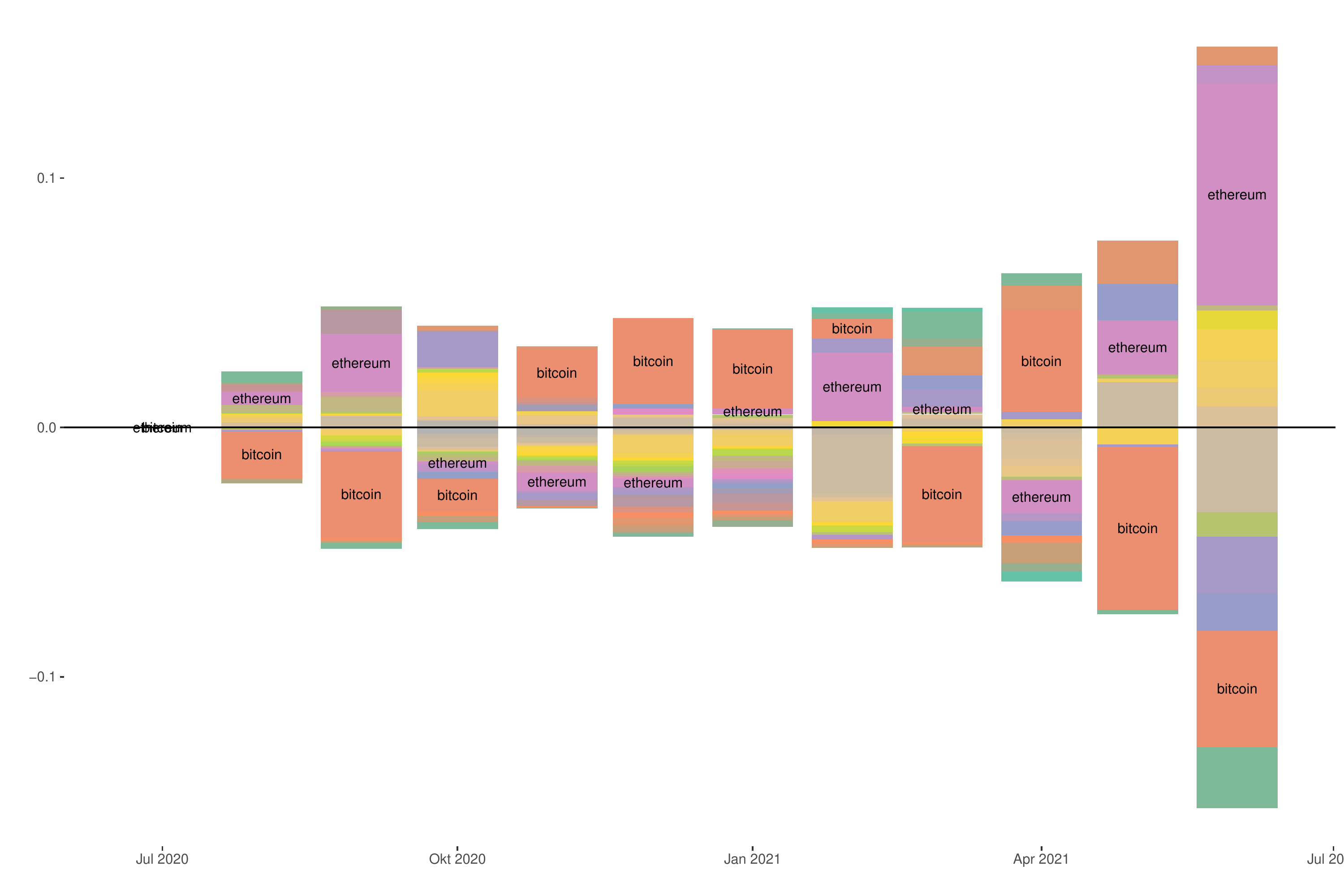}
\caption{Portion of the ETF portfolio that had to be rebalanced at the rebalancing dates. Each colour refers to a CC. Positive values correspond to a purchase/inclusion of CCs, negative values to the sale of CCs.
}
\label{graph:deltas_over_time}
\end{figure}

As shown in Figure \ref{graph::crix_constituents}, the number of index constituents varies from 5 to 30. Although the number of constituents varies greatly, this does not translate to the proportion of the ETF that needs to be rebalanced. Figure \ref{graph:deltas_over_time} shows the quantity of CCs (as a percentage of the total portfolio) that needed to be rebalanced on the rebalancing dates, i.e. the difference in the weight of each CRIX constituent between the current and previous rebalancing dates, we call it Delta: $\Delta_{cc,t} = weight_{cc,t} - weight_{cc,t-1}$. That is, positive values correspond to a purchase/inclusion of CC $cc$ at the current rebalancing date $t$, negative values correspond to the sale of CCs. Each color in the graph corresponds to a CC. To make the graph clear, the names of the CCs are not displayed, only those of Bitcoin and Ethereum. At the beginning of the analysis, all the capital was distributed among the constituents, so the sum of the $\Delta_{cc,0}$ is not zero but one. 
The share of the portfolio that had to be rebalanced varies between $2.2 \%$ (August 2020) and $15.2 \%$ (June 2021). On average, $5.2 \%$ of the portfolio had to be rebalanced. 

At first glance, there is no discernible correlation between the number of index constituents and the proportion of the portfolio that had to be rebalanced. Moreover, on the dates when the set of constituents is redetermined (1st week of Jan/Apr/Jul/Oct), it appears that no extraordinary portion of the portfolio needs to be reallocated. This suggests that it is not the number of constituents but rather the size of their weight that is important for the extent of rebalancing. As was already evident in the descriptive Figure \ref{graph:ETH_BTC}, a large part of the portfolio's weight is on BTC \& ETH, and this share is quite robust over time. The other CCs, no matter how many constituents the CRIX has, have significantly less weight. This leads to the observation that sometimes many currencies have to be sold/ bought, even though only small shares of the portfolio are rebalanced. There seems to be no direct correlation between the number of index constituents and the share of the portfolio that has to be rebalanced. 

The share of the portfolio that is updated varies between $2.2 \%$ (August 2020) and $15.2 \%$ (June 2021). These are not negligible amounts, especially when managing large portfolios. These quantities are the price one pays for accurate mapping of market dynamics. Although the portfolio only needs to be changed 5.2\% on average, the weightings of almost all constituents do change, resulting in up to 33 CCs (Oct \& Nov 2020)  needing to have their weightings updated. \\

\begin{figure}[h!]
    \centering
    \includegraphics[width=\textwidth]{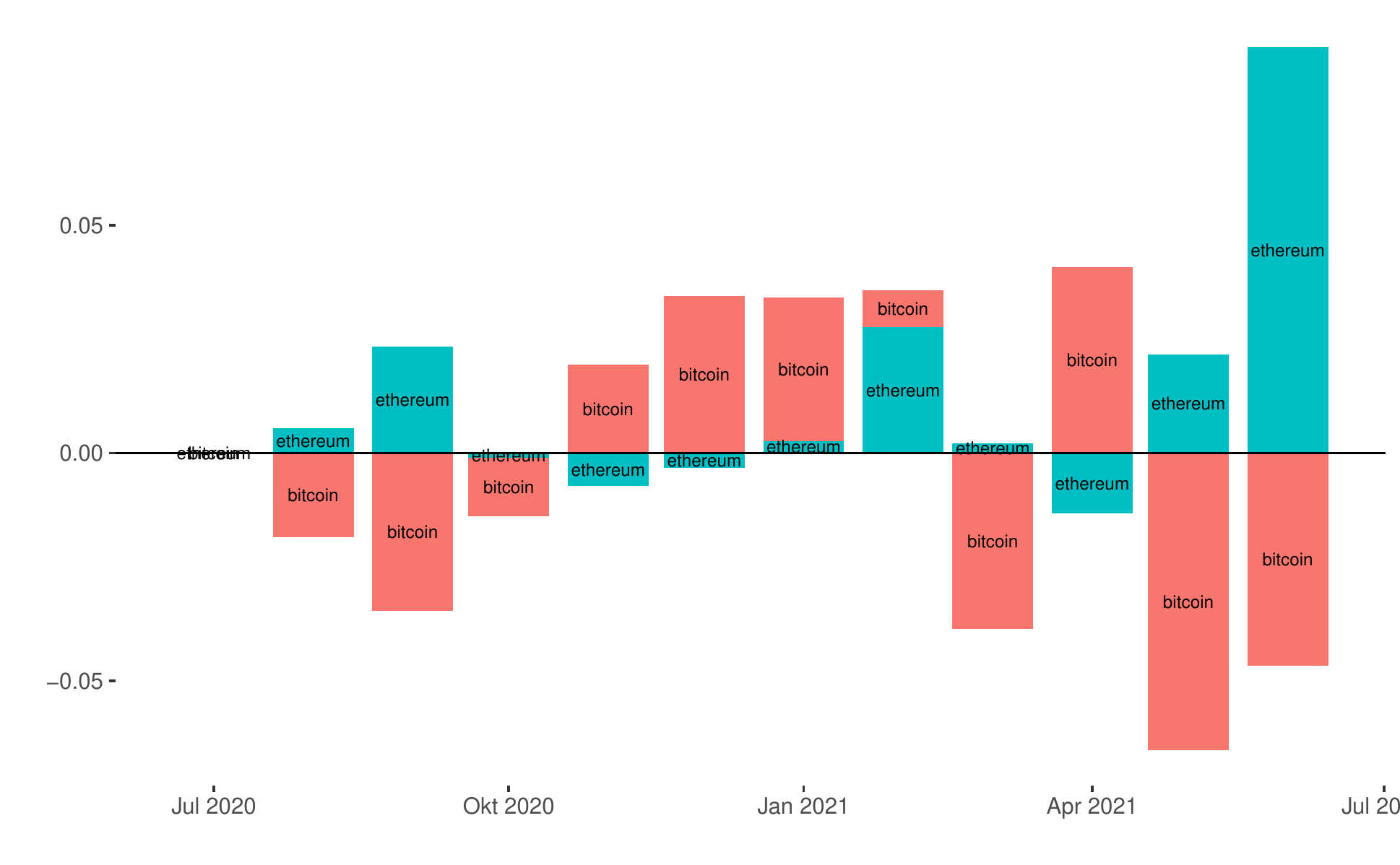}
    \caption{$\Delta_{cc,t}$ of Bitcoin and Ethereum over time. 
    }
    \label{graph:delta_btc_eth}
\end{figure}

The ETF constituents that have been rebalanced the most (i.e., that have the highest absolute $\sum_t|\Delta_{cc,t}|$) are Bitcoin ($\sum_t |\Delta_{BTC,t}| = 0.35$) and Ethereum ($\sum_t |\Delta_{ETH,t}| = 0.20$). Figure \ref{graph:delta_btc_eth} shows the $\Delta_{cc,t}$ of Bitcoin and Ethereum over time. One can see that on many rebalancing dates $\Delta_{BTC,t}$ and $\Delta_{ETH,t}$ have opposite signs, so when Bitcoin was sold, Ethereum had to be bought in. Furthermore, the (sale) purchase of a constituent is often followed by the opposite operation in the following rebalancing date. \\

\paragraph{
Costs} The issuance of a physical ETF involves the purchase and custody of the underlying assets. In our scenario analysis, we disregarded costs that are fixed and firm-specific, e.g., custody costs. Due to regular rebalancing of the CRIX weights and deposits/ withdrawals of capital, CCs need to be bought and sold. An analysis of the resulting trading fees and spreads is carried out in the next paragraphs. \\ 

Trading fees are very low in the CC sector, see Figure \ref{graph:fee_schedule}. Depending on the size of the transaction volume, the fees vary between $0.2\% - 0.5\%$. For the rising CC sector, it is rather interesting to study the spreads, especially for altcoins whose liquidity is lower than the liquidity of BTC and ETH (cf. \cite{brauneis2021}).  \\

\begin{figure}[h!]
    \centering
    \includegraphics[width=\textwidth]{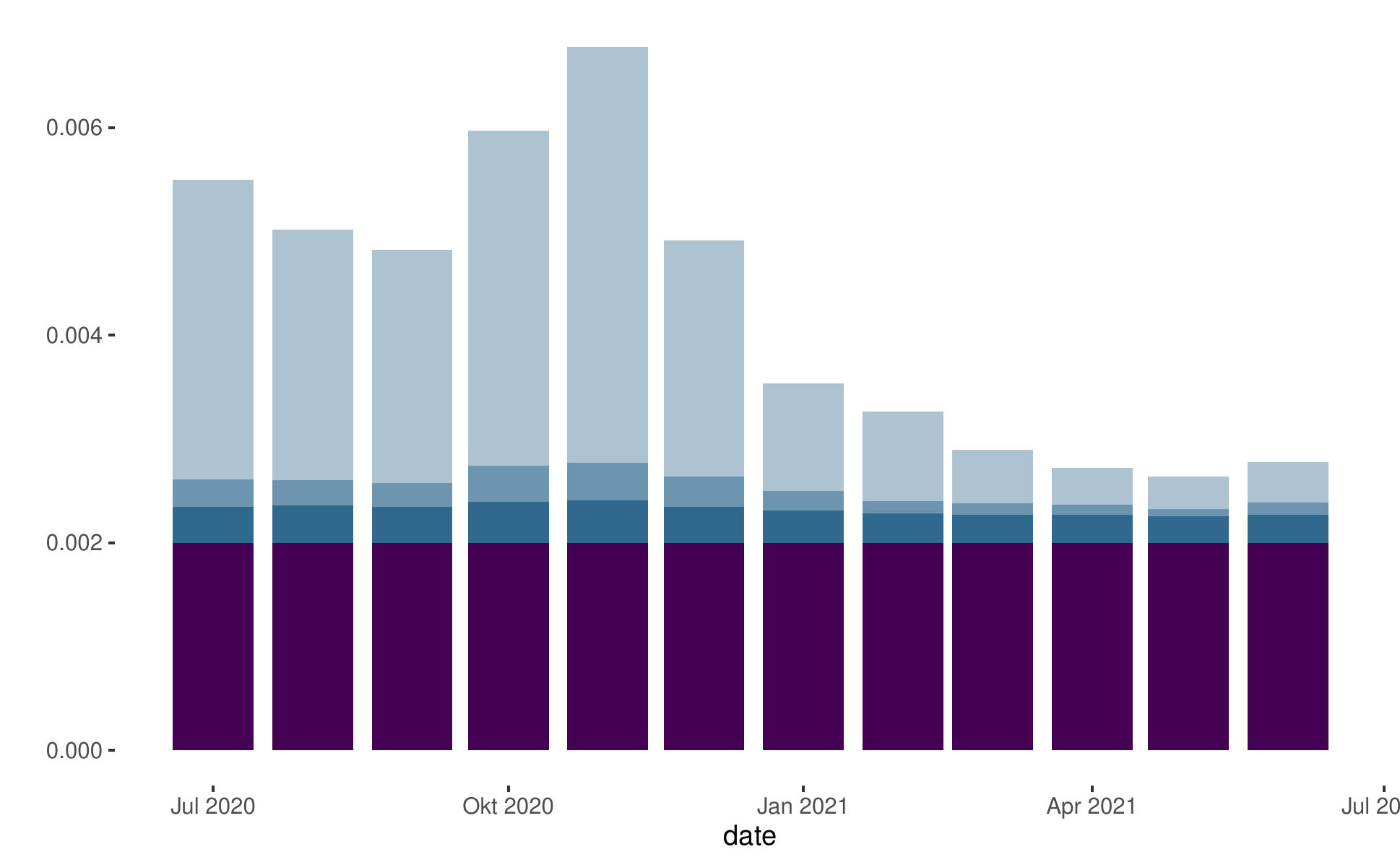}
    \caption{Stacked bar plot of \textcolor{trading_fees}{trading fees} and spreads (for \textcolor{a2}{$a=2$}, \textcolor{a5}{$a=5$}, \textcolor{a10}{$a=10$}, cf. Equation  \ref{eqn::relBTC}), weighted by $\Delta_{cc,t}$.
    }
    \label{graph:stacked_costs}
\end{figure}

In order to quantify the size of the transaction fees and spreads, we propose a scenario in which we start with an initial capital endowment of 1.000.000 USDT at 2020-07-01 and let the capital deposits grow as described in Section \ref{sec:costs_adjustment}. Figure \ref{graph:stacked_costs} presents a stacked bar plot of  spreads and fees for each rebalancing date. Fees and spreads are calculated as weighted average (by $\Delta_{cc,t}$) of the fees/spreads that incurred by the purchase/sale of the constituents at rebalancing.


As shown in Figure \ref{graph:stacked_costs}, spreads increase in 2020 until they peak in November 2020. During this period, spreads vary between 0.08\% and 0.5\% depending on the scaling factor $a$ (cf. Equation \ref{eqn::relBTC}). It is striking that spreads fall towards the end of the analysis period, after having peaked in November 2020. Based on the small sample of one year, it is difficult to derive a trend from this, however, this observation fits our expectations as we expect more liquid markets as the CC sector continues to consolidate. The increase in liquidity reduces the size of spreads and thus trading costs. 

\begin{figure}[h!]
\includegraphics[width=\textwidth]{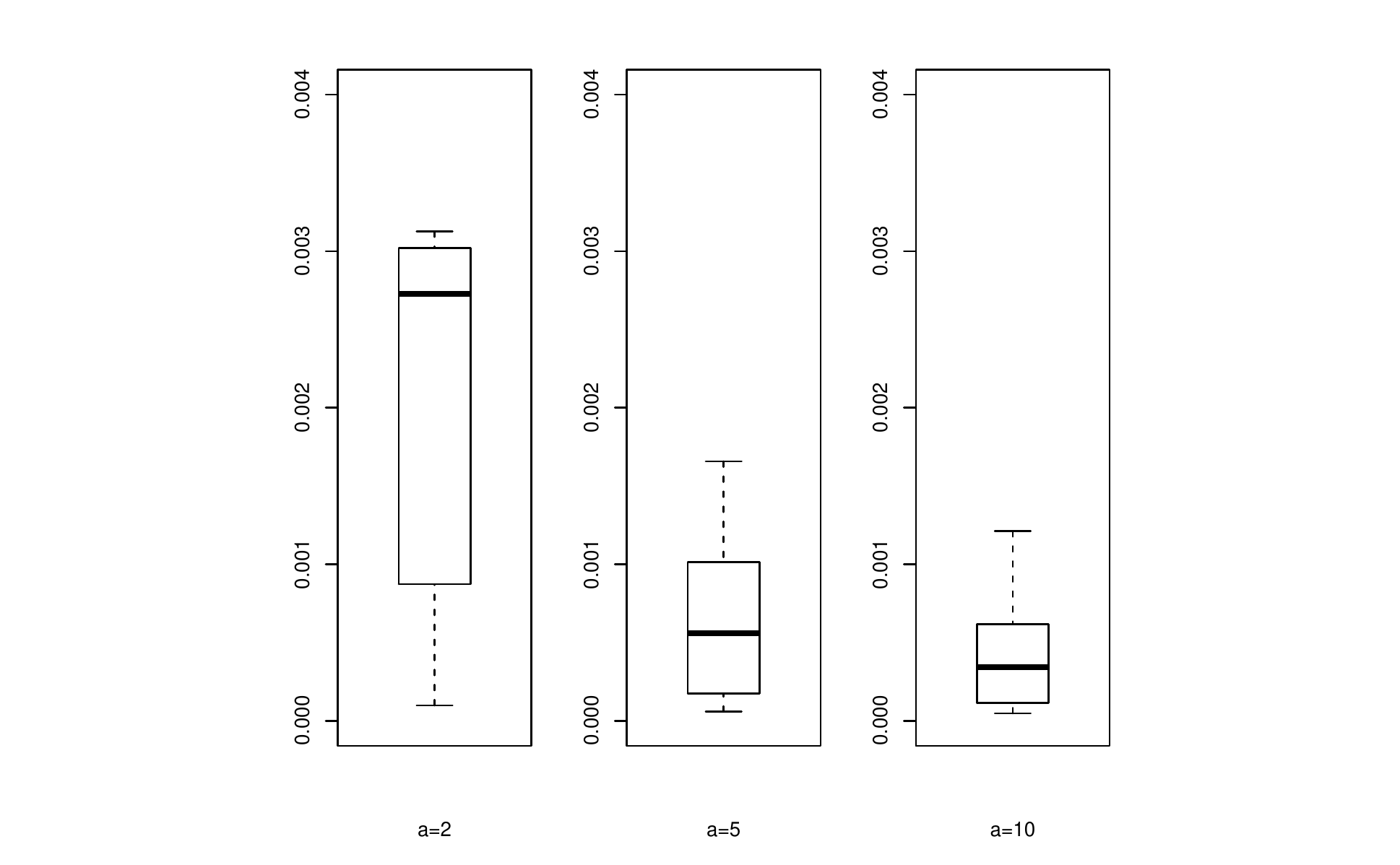}
\caption{Distributions of spreads that incurred at rebalancing, weighted by $\Delta_{cc,t}$ and given the scaling factors $a = \{2, 5, 10\}$
}
\label{graph:boxplot_costs}
\end{figure}

A more detailed overview of the distribution of the spreads is provided by the boxplots in Figure \ref{graph:boxplot_costs}. The three boxplots depict the distribution of spreads, weighted by their size $\Delta_{cc,t}$. The median of the spreads, depending on the assumption about the scaling factor $a$, varies for the analysis period between 0.03\% and 0.27\%. 

The size of the spreads depends on the size of the transactions. As stated in Section \ref{subsection::spreads}, we assume a linear relationship between transaction volume and spread size. However, the rare presence of high trade volume data points means that the analysis is not robust at the upper tail.  This implies that inference regarding the spreads of large transactions is not meaningful. In extreme cases, large transactions may even move the market. Such a scenario is not unlikely, since the in- \& exclusion of a constituent can lead to large reallocations of capital if the volume of the ETF is large. 

\begin{figure}[h!]
\includegraphics[width=0.9\textwidth]{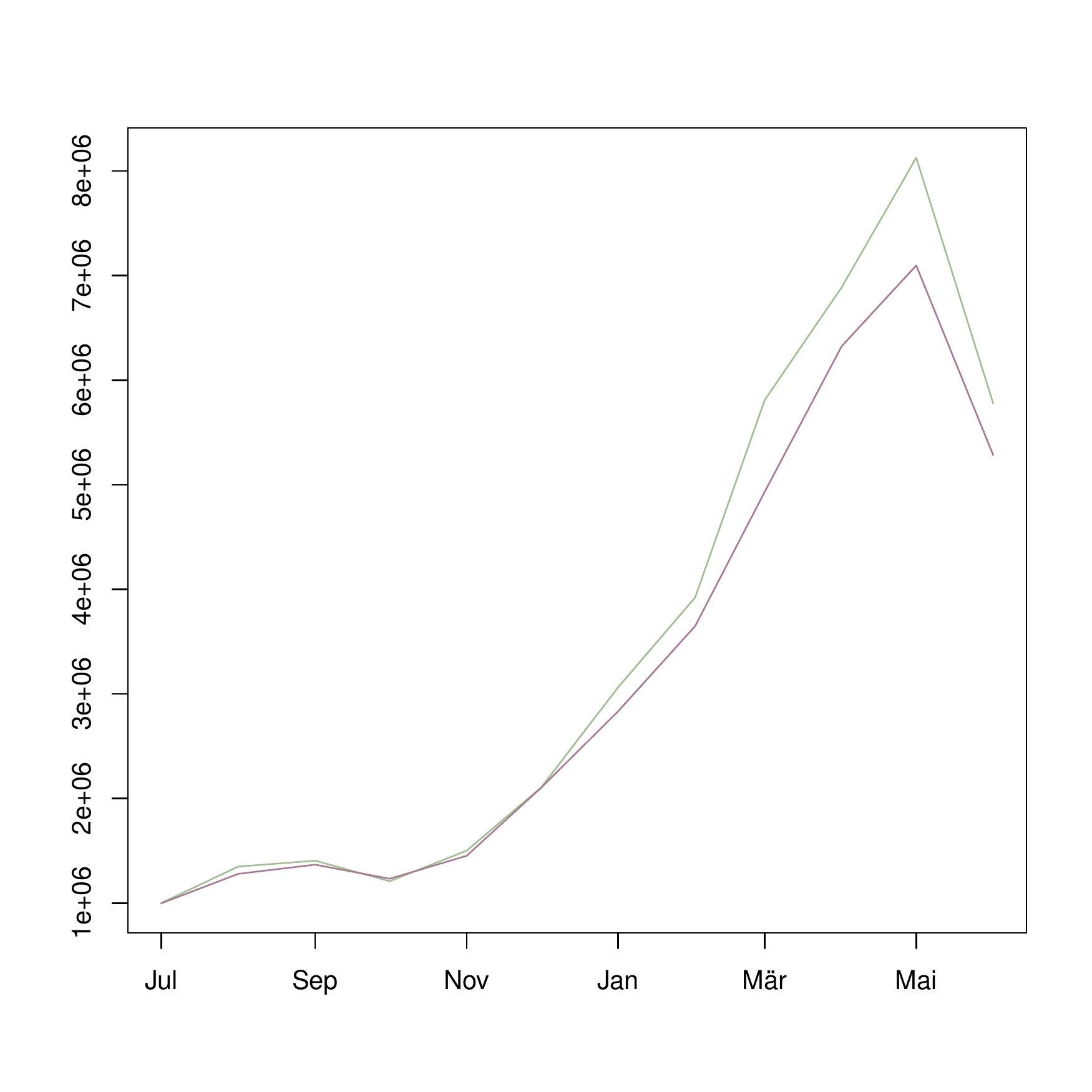}
\caption{Value of the {\color{etf}ETF portfolio} and the {\color{crix}CRIX}, scaled to 1e06 July 2020 and inflated by the capital flows as described in Section \ref{sec:costs_adjustment}.
}
\label{graph:crix_etf}
\end{figure}

\paragraph{
ETF portfolio value and CRIX} To make sure that everything went right during the rebalancing procedure, we compare the value of the ETF portfolio with the CRIX, inflated by the capital flows as described in Section \ref{sec:costs_adjustment}. \newpage Figure \ref{graph:crix_etf} illustrates the value of the ETF portfolio (dark pink line) and the CRIX (green line). The values of both time series grow, because the CC sector experienced a massive growth during the period of analysis and in addition, the deposits grow as described in Section \ref{sec:costs_adjustment}. The value of the ETF portfolio is always slightly below the CRIX due to trading fees and spreads. As rebalancing takes place on a regular basis, the difference between the two time series increases over time. 

\begin{figure}[h!]
    \centering
    \includegraphics[width=0.9\textwidth]{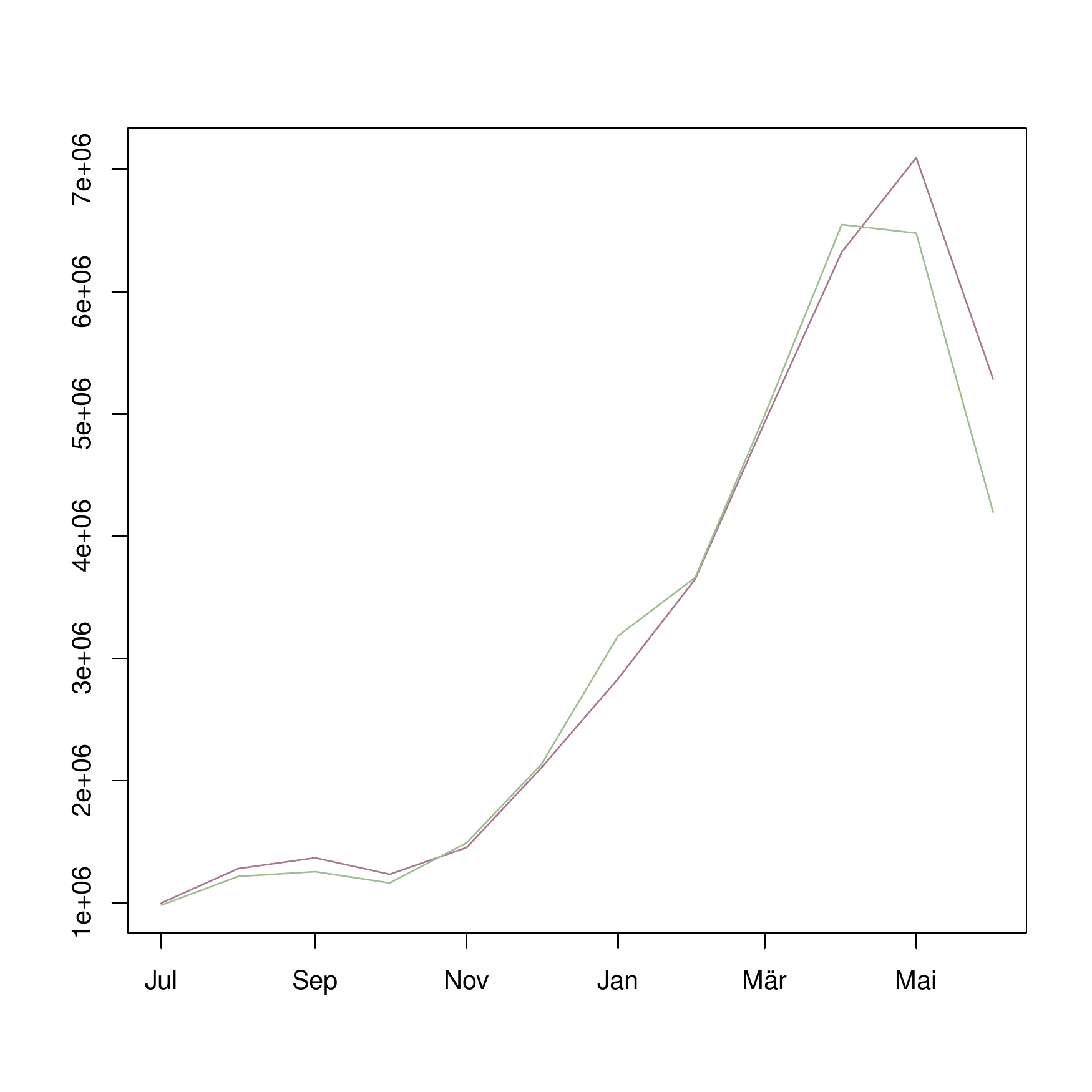}
    \caption{Time series of {\color{crix}BTC}, scaled to 1e06 in July 2020 and inflated by capital deposits as described in Section \ref{sec:costs_adjustment}, and the value of the {\color{etf}ETF portfolio}.}
    \label{graph:btc_etf}
\end{figure}

\paragraph{Performance} The performance of the CRIX ETF is difficult to assess, as the CC sector is highly non-stationary and thus the performance depends on the time interval under review. As a benchmark, we compare the performance of the CRIX ETF with a simple investment in BTC. Figure \ref{graph:btc_etf} displays the value of the ETF portfolio (pink line) and BTC (black line). Both time series have similar dynamics, but most of the ETF's values are slightly lower than those of BTC, which means that an investment in Bitcoin is usually more profitable. Only at the end of the analysis period the value of Bitcoin collapses earlier than that of the ETF. This observation is consistent with \cite{nakagawa2022}, who report weaker correlation of Bitcoin and Ethereum during periods of high market uncertainty. However, the Sharpe ratios contradict the first impression, the ETF (0.89) outperforms BTC (0.66). This is probably due to the bearish market dynamics from May 2021 onwards; the ETF seems to be able to cushion the slump of BTC, the risk diversification works. 

\paragraph{
The volatility of the CC sector and resulting consequences for the CRIX ETF} 
The CC sector is in constant transformation, which is reflected in the high volatility and the emergence and disappearance of CCs. How do these dynamics affect the CRIX ETF? In particular, how does it affect the number and composition of constituents, and thus the rebalancing frequency and costs? \\

The volatility of the entire CC sector is partly reflected in the composition of the ETF: during the period of analysis, adjustments had to be made to the ETF portfolio at each rebalancing date, both the weights and the composition have changed frequently. However, the biggest portion of the portfolio (85\% -98\%) remains unaffected by the rebalancing, "only" a 2\% - 15\% share has to be changed. This is mainly because the CC sector is strongly dominated by Bitcoin and Ethereum, and their share has remained quite robust over time (together they represent about 80\% of the ETF portfolio). This has the disadvantage of losing risk reduction through diversification, which is one of the main reasons for using ETFs as a financial instrument. At the same time, however, BTC and ETH are the CCs that have had to be rebalanced the most. Overall, two effects can be observed: on the one hand, smaller CCs (with low weighting) are frequently added/sold in the ETF, and on the other hand, the largest CCs are subject to the largest reallocations. \\

This result is interesting as it shows that the CRIX, as an adaptive index, adapts to larger fluctuations of the total market, but remains robust in its core. The reason for this lies in the construction of the CRIX: starting with the largest CC by market capitalization, additional CCs are added to the CRIX portfolio so that it is as close as possible to the total market. We suspect that one explanation for the robustness of the CRIX lies in the cointegration relationships of the CCs: as \cite{keilbar2021} have shown, there are cointegration relationships between the top CCs by market capitalization. If the price dynamics of the CCs are similar, the inclusion of more CCs to the index does not lead to an information gain, and thus, a smaller set of CCs is needed to map the market dynamics. \\

From the perspective of the issuer of the ETF, it is not optimal that on average 5.2\% of the portfolio has to be changed on each rebalancing date. However, this is the price one pays for an accurate mapping of the entire CC sector. It is advantageous that the largest adjustments are to be made to the most liquid CCs, as this limits the costs. \\

In general, statements and findings about the CC sector should be taken with caution, as the dynamics of this sector are highly non-stationary and thus all statements are time depending.

\section{Discussion of Results}

This section begins with a discussion of the robustness of the estimation results, followed by a discussion of the implications of the results for the practical implementation of a CRIX ETF.

\paragraph{Robustness of Estimation Results} Large parts of the analysis are deterministic in nature, as the values of the CRIX and the weights of its components were taken as given. The uncertainty mainly concerns the structure of spreads and the modeling of capital deposits/ withdrawals. Both points are closely related, as they affect/ are affected by the capital volume the ETF manages. In principle, the investment volume can be scaled up as desired. This does not change the nature and results in terms of quantities that need to be rebalanced, as all values in the analysis were expressed as percentages. Trading fees are also robust, as the structure of the fee schedule is deterministic. However, it becomes more difficult when it comes to the size of the spreads and any market movements. Placing large positions can lead to market movements that are difficult to assess historically. While one could simulate the market movements of all CCs at all historical rebalancing dates based on the distribution of their returns, this would result in very wide confidence intervals as the variance of any market movement on each date multiplies over time.


\paragraph{Discussion of the practical Implications of the Results} The results of the scenario analysis have far-reaching consequences for the practical implementation of such an ETF, especially with respect to risk diversification and optimization of rebalancing quantities. One of the basic ideas of ETFs is to reduce the risks of individual securities through diversification. In the case of the CRIX ETF, diversification is not given as the weights of Bitcoin and Ethereum dominate the portfolio and their weights are robust over the period of analysis. In general, diversification in the CC sector is complicated because \cite{keilbar2021} identified cointegration relationships among the top CCs by market capitalization. However, whether these cointegration relationships will endure in the future is not certain. There are several ways to reduce the weights of Bitcoin and Ethereum. Since the CRIX weights its individual constituents according to their market capitalization, it would be possible to take the logarithmized market capitalization instead, for example. Also, one could simply introduce a cap/floor for the weights of the individual constituents, and divide the remaining weights among the other constituents. \\

The second concern tackles the rebalancing quantities $\sum_t \Delta_{cc,t}$. The constituents with the largest rebalancing share over the period of analysis are Bitcoin and Ethereum, and as it can be seen in  Figure \ref{graph:delta_btc_eth}, their deltas often have opposite signs (i.e. if Bitcoin is sold, Ethereum must be bought, and vice versa), and often the (sale) purchase of a constituent is followed by the opposite operation in the next rebalancing period (e.g. Bitcoin is bought at time $t$ and sold at time $t+1$, etc.). We propose to modify the CRIX ETF algorithm so that the transitions at the rebalancing dates are smoother, i.e., the quantities $\sum_t \Delta_{cc,t}$ are minimized. In the practical implementation of an ETF, such an optimization would reduce trading costs. 

\section{Conclusion}\label{sec::conclusion}
Investments in digital assets remain risky due to high volatility. An ETF on an CC index can be a tool to benefit from the gains of this emerging sector while reducing risks through diversification. In our scenario analysis, we have gained interesting insights into the operation, risks and costs of issuing a crypto ETF. In particular, the analysis produced the following findings:  \\

As for the rebalancing mechanism, the analysis showed that the average share of the portfolio that had to be rebalanced was 5.2\% during the analysis period. In addition, much of the weight of the CRIX ETF is distributed between BTC and ETH (between 70\% - 80\%). On the one hand, this leads to the core of the ETF remaining robust, but at the same time it reduces the diversification among the CCs. \\

Regarding the cost structure of issuing a CRIX ETF, the results show that i) the cost of trading CCs is low (assuming the fee schedule of Coinbase), ii) spreads decrease over the analysis period and vary (depending on the assumptions about the structure of spreads and transaction volume) on average between 0.03\% and 0.27\%. \\

Regarding the risks involved in managing a CRIX ETF: since the CRIX is a dynamic index that adjusts to overall market dynamics, a portion of the portfolio must be reallocated on a regular basis. This part averages 5.2\%, and this is the price one pays for an accurate mapping of the overall market dynamics of this volatile sector. If the ETF has a high volume, 5.2\% is not negligible. 
\\

Furthermore, we have some recommendations: first, to guarantee the basic idea of risk diversification of an ETF, we recommend limiting the dominance of BTC and ETH. Second, it is useful to modify the CRIX ETF algorithm to include minimization of rebalancing quantities through a constraint to make the ETF's weights more robust and to limit trading costs. \\

So far, we constructed a physical ETF on the CRIX. Future work can focus on the synthetic ETFs constructed by derivatives. There will certainly be liquid markets for derivatives on smaller CCs in the near future, paving the way for synthetic ETFs. \\

\section*{List of Abbreviations}

\begin{alignat*}{3}
   &\textbf{BTC}  &  &\dots \text{Bitcoin} \\
   &\textbf{CC}   &  &\dots \text{Cryptocurrency} \\
   &\textbf{ETF}  &  &\dots \text{Exchange Traded Fund} \\
   &\textbf{ETH}  &  &\dots \text{Ethereum} 
\end{alignat*}

\bibliography{Extract_ETF/references}

\begin{thebibliography}{13}
\providecommand{\natexlab}[1]{#1}
\providecommand{\url}[1]{\texttt{#1}}
\expandafter\ifx\csname urlstyle\endcsname\relax
  \providecommand{\doi}[1]{doi: #1}\else
  \providecommand{\doi}{doi: \begingroup \urlstyle{rm}\Url}\fi

\bibitem[Amihud(2002)]{amihud2002}
Y.~Amihud.
\newblock Illiquidity and stock returns: cross-section and time-series effects.
\newblock \emph{Journal of financial markets}, 5\penalty0 (1):\penalty0 31--56,
  2002.

\bibitem[Brauneis et~al.(2021)Brauneis, Mestel, and Theissen]{brauneis2021}
A.~Brauneis, R.~Mestel, and E.~Theissen.
\newblock What drives the liquidity of cryptocurrencies? a long-term analysis.
\newblock \emph{Finance Research Letters}, 39:\penalty0 101537, 2021.

\bibitem[Burda(2021)]{burda2021}
M.~C. Burda.
\newblock Valuing cryptocurrencies: Three easy pieces.
\newblock \emph{IRTG 1792 Discussion Paper}, 2021.

\bibitem[Dyhrberg et~al.(2018)Dyhrberg, Foley, and
  Svec]{dyhrberg2018investible}
A.~H. Dyhrberg, S.~Foley, and J.~Svec.
\newblock How investible is bitcoin? analyzing the liquidity and transaction
  costs of bitcoin markets.
\newblock \emph{Economics Letters}, 171:\penalty0 140--143, 2018.

\bibitem[Grobys et~al.(2020)Grobys, Dufitinema, and Sapkota]{grobys2020}
K.~Grobys, J.~Dufitinema, and N.~Sapkota.
\newblock On the tail risk of cyberattacks in the bitcoin market.
\newblock \emph{Available at SSRN 3733810}, 2020.

\bibitem[H{\"a}rdle and Trimborn(2015)]{davis2016}
W.~K. H{\"a}rdle and S.~Trimborn.
\newblock Crix or evaluating blockchain based currencies.
\newblock \emph{Oberwolfach Reports „The Mathematics and Statistics of
  Quantitative Risk“}, \penalty0 (42/2015), 2015.

\bibitem[H{\"a}rdle et~al.(2020)H{\"a}rdle, Harvey, and Reule]{haerdle2020}
W.~K. H{\"a}rdle, C.~R. Harvey, and R.~C. Reule.
\newblock Understanding cryptocurrencies.
\newblock \emph{Journal of Financial Econometrics}, 18:\penalty0 181--208,
  2020.

\bibitem[H{\"a}usler and Xia(2022)]{haeusler2022}
K.~H{\"a}usler and H.~Xia.
\newblock Indices on cryptocurrencies: an evaluation.
\newblock \emph{Digital Finance}, 4:\penalty0 149–167, 2022.

\bibitem[Keilbar and Zhang(2021)]{keilbar2021}
G.~Keilbar and Y.~Zhang.
\newblock On cointegration and cryptocurrency dynamics.
\newblock \emph{Digital Finance}, 3:\penalty0 1--23, 2021.

\bibitem[Lee et~al.(1993)Lee, Mucklow, and Ready]{lee1993spreads}
C.~M. Lee, B.~Mucklow, and M.~J. Ready.
\newblock Spreads, depths, and the impact of earnings information: An intraday
  analysis.
\newblock \emph{The Review of Financial Studies}, 6\penalty0 (2):\penalty0
  345--374, 1993.

\bibitem[Nakagawa and Sakemoto(2022)]{nakagawa2022}
K.~Nakagawa and R.~Sakemoto.
\newblock Market uncertainty and correlation between bitcoin and ether.
\newblock \emph{Available at SSRN}, 2022.

\bibitem[Trimborn and H{\"a}rdle(2018)]{trimborn2018}
S.~Trimborn and W.~K. H{\"a}rdle.
\newblock Crix an index for cryptocurrencies.
\newblock \emph{Journal of Empirical Finance}, 49:\penalty0 107--122, 2018.

\bibitem[Wang and Yau(2000)]{wang2000trading}
G.~H. Wang and J.~Yau.
\newblock Trading volume, bid--ask spread, and price volatility in futures
  markets.
\newblock \emph{Journal of Futures Markets: Futures, Options, and Other
  Derivative Products}, 20\penalty0 (10):\penalty0 943--970, 2000.

\end{thebibliography}

\end{document}